# Assessing Climate Vulnerability Risk for Substations in Massachusetts Via Sensitivity Analysis

Hritik Gopal Shah  
Resiliency and Reliability Planning  
Eversource Energy, Westwood, MA, USA  
hritik.shah@eversource.com

Elli Ntakou  
Resiliency and Reliability Planning  
Eversource Energy, Westwood, MA, USA  
elli.ntakou@eversource.com

*Abstract*— The electric grid is increasingly vital, supporting essential services such as healthcare, heating and cooling transportation, telecommunications, and water systems. This growing dependence on reliable power underscores the need for enhanced grid resilience. This study presents Eversource's Climate Vulnerability Assessment (CVA) for bulk distribution substations in Massachusetts, evaluating risks from storm surge, sea level rise, precipitation, and extreme temperatures. The focus is on developing a cost-efficient model to guide targeted resilience investments. This is achieved by overcoming the limitations of single-variable analyses through hazard-specific assessments that integrate spatial, climate, electrical asset, and other relevant data; and applying sensitivity analysis to establish data-driven thresholds for actionable climate risks. By integrating geospatial analysis and data modeling with power engineering principles, this study provides a practical and replicable framework for equitable, data-informed climate adaptation planning. The results indicate that thresholds for certain climate hazards can be highly sensitive and result in significantly larger sets of stations requiring mitigation measures to adequately adapt to climate change, indicating that high-fidelity long-term climate projections are critical.

## I. Introduction and Motivation

As climate patterns shift, extreme weather events are becoming more frequent and intense, diverging from historical norms. While climate impacts are often framed by long-term average temperature increases, growing temperature variability is driving more severe weather. These changes pose significant risks to the power sector, threatening infrastructure, disrupting operations, and altering energy supply and demand [1]. Key stressors include rising temperatures, wildfires, severe storms, and sea level rise require utilities to adopt flexible, forward-looking strategies focused on proactive resilience.

Eversource is a non-vertically integrated utility serving approximately 4.4M customer meters across the states of New Hampshire, Massachusetts and Connecticut and is committed to providing safe and reliable service to all its customers. [2] presents an analysis of the number of outages Eversource incurred as part of storms in the 1980-2020 window. It concludes that climate intensification is already being observed, since the data clearly indicate an increasing intensity trend with more recent storms resulting in a higher number of outages. The aforementioned climate-related risks incurred at the Eversource territory are compounded by aging infrastructure and other challenges like increasing DER penetration levels and electrification-based loads [3], [4].

In response to increasing climate risks, several institutions have developed tools to help utilities assess vulnerabilities and improve resilience. The U.S. Department of Energy (DOE), Pacific Northwest National Laboratory (PNNL), and Electric Power Research Institute (EPRI) [5] [6] [7] offer frameworks and guidance for evaluating asset risk and planning mitigation strategies. While these resources offer valuable high-level guidance, they often rely on generalized strategies that may not translate effectively across diverse geographic contexts. Regional differences in topography, infrastructure, population density, and localized climate projections significantly influence climate vulnerabilities.

[8] offers a comprehensive review of climate change impacts on the bulk power system; however, it lacks a detailed sensitivity analysis at the component level for individual power system assets. Similarly, while [9] is focused on heat events and the vulnerabilities of electricity infrastructure to such events in Los Angeles County. [10] provides a comprehensive review of how extreme temperature events, such as heat domes, polar vortices, and icing affect T&D grid equipment.

Across the United States, several major utilities—including Duke Energy, National Grid, Southern California Edison, and Con Edison [11], [12] and [13] have published climate vulnerability assessments, either submitted to regulatory bodies or made publicly available through their websites. These reports typically present projections of asset exposure to climate hazards such as extreme heat, flooding, and sea level rise for future timeframes ranging from 2030 to 2080.

However, a common limitation among industry and academic climate vulnerability assessment is the lack of data-based analytics to connect assets' exposure to actionable items and prioritizing mitigations. Such quantitative approaches would enable repeatable climate analyses, across different geographic regions, with different climatic conditions and relevant climate hazards. Further, such quantitative approaches can be edited to accommodate utility territories of different sizes.

This paper is part of Eversource's data driven plan to increase resilience and hedge against climate hazards. Specifically, the paper focuses on the exposure of bulk substations within Eversource's Massachusetts territory to climate risks. The focus is on the bulk substations and their assets, because it is both costly and time consuming to replace or upgrade bulk substations, while at the same time, failure of substation equipment can result in extended interruptions. Because the application is Massachusetts, the relevant climate hazards for stations are (i) sea level rise, (ii) storm surge, (iii) precipitation and (iv) extreme temperatures.

This paper's contributions are centered around the operationalization of the climate projection inputs to the standard Company planning and operations tools and the sensitivity analysis to create actionable thresholds, optimize mitigations at the asset level and prioritize mitigative work. In particular, this paper's contributions include:

- Integrating the climate hazard projections and flagged high-risk assets into the Company's Geographic Information Systems (GIS) to enable climate-risk based decision making and integrated planning solutions.

- Developing a novel Geospatial Framework for Climate Hazard Vulnerability Assessment. A novel ArcGIS-based methodology integrates downscaled 2050 climate projections to identify vulnerable assets at the substation level, offering a Massachusetts-specific alternative to broad national models.

- Conducted multi-variable sensitivity analyses to define actionable thresholds for climate hazards, enabling targeted, cost-efficient resilience investments and optimized resource allocation.

## II. Climate Vulneribility Assesment in Massachusetts

Following U.S. Department of Energy (DOE) guidance, this study defines risk as the product of threat, vulnerability, and consequence [1]. Threats are the probabilistic projections of climate hazards, namely temperature extremes, precipitation, sea level rise, and storm surge. Vulnerability reflects asset susceptibility, while consequence measures the impact of failure of an asset on energy service.

Because the focus on this analysis is climate risk of bulk substations in Eversource's Massachusetts territory; (i) sea level rise, (ii) storm surge, (iii) precipitation and (iv) temperature extremes, were the four hazards that were identified as most relevant. Temperature and precipitation projections were based on downscaled Shared Socioeconomic Pathways (SSP) 2-4.5 (GHG emission stabilization) scenario for 2050, while sea level rise and storm surge data were sourced from NOAA. Coastal substations are most vulnerable to sea level rise and storm surge, while stations are more susceptible to inland flooding. Substation transformers are the station asset that is more at risk from rising temperatures.

### A. Exposure of Bulk Substations to Sea Level Rise

This study evaluates the potential impact of projected sea level rise (SLR) on bulk substations within Eversource's Massachusetts territory, using a 2-foot sea-level rise (SLR) scenario, as projected by the National Oceanic and Atmospheric Administration (NOAA) [14]. We use the 2 feet SLR scenario as it aligns with the projected temperature levels under SSP 2-4.5 by 2050. The methodology treats sea level rise (SLR) as a chronic hazard, identifying substations not only in projected inundation zones but also those with significantly reduced distance to water under a 2-foot SLR scenario, ensuring early, proactive risk detection. The assessment follows a four-step geospatial and analytical process:

1. Ingesting NOAA SLR projections as a layer in the Company's ArcGIS systems

2. Baseline Distance Calculation

The present-day shortest distance from each substation $S_i$ to the nearest coastal water body was computed using Euclidean distance (1):

$$d_0(S_i) = \min_{x \in C_0} \|\text{Location}(S_i) - x\| \quad (1)$$

Where, $S_i$ is $i$-th substation, $C_0$ denotes current coastline and $d_0(S_i)$ is the present-day Euclidean distance from $S_i$ to $C_0$.

3. Future Projected Distance under 2-ft SLR

The projected coastline under a 2-ft SLR scenario was used to recalculate the distance (2) from each substation to the encroached shoreline under the 2-ft SLR scenario.

$$d_{SLR}(S_i) = \min_{x \in C_{2ft}} \|\text{Location}(S_i) - x\| \quad (2)$$

Where, $C_{2ft}$ is the projected coastline under 2-ft SLR and $d_{SLR}(S_i)$ is the projected Euclidean distance from $S_i$ to $C_{2ft}$.

4. Station Flagging Criteria

Substations were flagged for further review if (2) the projected distance to the shoreline is zero, indicating that the station is in an area of projected coastal encroachment, or the inland movement of the coastline exceeds a threshold proportion of the original distance. Total change in distance is given by $\Delta d(S_i) = d_0(S_i) - d_{SLR}(S_i)$. Then flagged station because of inland flooding is denoted by a binary indicator function $I_{flagged}(S_i)$ (3).

$$I_{flagged}(S_i) = \begin{cases} 1, \text{if } d_{SLR}(S_i) = 0 \mid \frac{\Delta d(S_i)}{d_0(S_i)} \geq \varphi \\ 0, \text{otherwise} \end{cases} \quad (3)$$

Where $\varphi$ is the percent change of the station's distance to the water and the parameter over which sensitivity analysis is performed, see section III.

It is important to note that projected flooding at a substation location does not necessarily imply functional inundation of equipment. Sea level rise levels are projections with inherent uncertainty, and substations are constructed with elevation buffers that may already provide adequate protection.

### B. Exposure of Bulk Stations to Storm Surge

This section evaluates the vulnerability of bulk substations to coastal flooding from hurricane-induced storm surge using NOAA's Sea, Lake, and Overland Surges from Hurricanes (SLOSH) model [15], which provides inundation projections for various hurricane categories. While Category 3 hurricanes are more intense than Category 1 or 2, they are rare in Massachusetts; only two have occurred since 1842, compared to nine Category 1 storms [16]. Given this historical frequency, the analysis focuses on Category 1 hurricanes as the most probable threat. This risk-based approach ensures that mitigation strategies align with likely scenarios, supporting efficient resource allocation. The assessment is implemented through a GIS-based automation framework using ArcGIS and Python, enabling high-resolution spatial analysis of substation exposure.

1. SLOSH Data Integration

NOAA's SLOSH model outputs were ingested into the GIS system. For each substation $S_i$, the projected storm surge water depth $h_{storm}(S_i)$ at the substation location under a Category 1 hurricane scenario was extracted. Substations are flagged if

$$h_{storm}(S_i) > 0 \tag{4}$$

2. FEMA Base Flood Elevation (BFE) Data Integration & Processing

Utilities have traditionally relied on FEMA's Base Flood Elevation (BFE) maps to assess substation flooding risk. These maps provide projected floodwater levels relative to the North American Vertical Datum of 1988 (NAVD88). To evaluate flood vulnerability, the BFE at a given substation location $S_i$ be $BFE(S_i)$., is compared against the ground elevation at that location, $e_{ground}(S_i)$ [17], which is derived from Digital Elevation Model (DEM) data—also referenced to NAVD88. A substation is flagged as flood-prone if the BFE exceeds the ground elevation at that site.

$$BFE(S_i) > e_{ground}(S_i) \tag{5}$$

3. Storm Surge Risk Flagging

Substations that are either flagged by (4) or (5) are considered as exposed to Storm. Potential mitigation strategies include elevation retrofits, flood barriers, waterproofing of critical equipment, and updates to design standards.

C. Exposure of Substations to Inland Flooding

While coastal substations are at risk from sea level rise and storm surge, inland substations are susceptible to flooding from extreme precipitation. Global climate science models, like those used for the SSP scenarios of this climate study, typically use variables like 5-day total precipitation to quantify precipitation. In our study, projections showed a maximum of 6.5 inches over five days and 3.9 inches in a single day for Massachusetts. However, these metrics fail to capture flash flood events, which occur over a few hours. Historical data strongly indicate this shortcoming. For example, New Bedford saw 9.5 inches of rain in a few hours in 2021, and Leominster and Westford recorded nearly 10 inches in six hours in 2023 [18]. These short-duration events are key drivers of severe inland flooding and require higher-resolution modeling for effective mitigation.

To address this, we propose a geospatial methodology to evaluate inland bulk substation exposure to pluvial (surface water) and fluvial (riverine) flooding [19]. The analysis uses ArcGIS and Python-based automation, incorporating high-resolution elevation data from the USGS Digital Elevation Model (DEM) [17] and hydrography layers.

1. Pluvial Flooding Risk – Local Elevation Analysis

To assess pluvial flooding risk, a spatial analysis was developed to determine whether the station is a local low elevation area, making it a potential collection point for surface water during heavy rainfall. The ground elevation at the substation location, $e(S_i)$ was extracted from the USGS DEM [17]. Additionally, ten random reference points $\{P_{i1}, P_{i2}, P_{i3}, \ldots, P_{i10}\}$ were generated within $r_s$ mile radius from each substation location, and their elevations $\{e(P_{ij})\}$ were also extracted. $r_s$ represents the radius in miles, within which the 10 comparison points are generated. Sensitivity analysis performed in III to evaluate the impact of varying $r_s$.

A substation is flagged for pluvial flooding if

$$I_{pluvial}(S_i) = \begin{cases} 1, \text{ if } e(S_i) = \min(\{e(S_i), e(P_{i1}), \ldots, e(P_{i10})\}) \\ 0, otherwise \end{cases} \tag{6}$$

2. Fluvial Flooding Risk – Proximity to Water Bodies

To evaluate fluvial flooding risk, the distance from each substation $S_i$ to the nearest inland water body $W_k$ (e.g., river, lake, or pond) was calculated using GIS spatial analysis. A substation is considered at risk if

$$\min_k \|Location(S_i) - Location(W_k)\| \leq \delta \tag{7}$$

Where $\delta$ represents the distance of the station to a local inland body of water . Sensitivity analysis performed in III to evaluate the impact of varying $\delta$ to assess which rising water levels may pose a flooding threat.

3. Final Substation List

Substations that meet both criteria (6) and (7) are flagged as high-risk for inland flooding. This integrated approach ensures that substations vulnerable to both surface water accumulation and riverine overflow are prioritized for further engineering review and mitigation planning.

D. Exposure of Station Transfomer to Warming Temperature

Transformer sizing and operational ratings are closely tied to ambient temperature conditions, with seasonal assumptions defined in utility standards; summer normal (25°C), summer emergency (40°C), winter normal (0°C), and winter emergency (10°C) aligned with ANSI loading limits. In a warming climate, these assumptions may no longer hold, prompting a reevaluation of transformer loading practices. Transformer thermal aging is governed by the hottest spot temperature (THST), influenced by ambient temperature, loading, and internal thermal inertia. Sustained high THST accelerates insulation degradation, especially during heat waves when elevated temperatures coincide with peak demand [20].

The projected climate data for 2050 includes projections of daily maximum temperatures and the number of two-day heat wave events, where a heat wave is defined as consecutive days exceeding the 90th percentile of historical maximum temperatures. This analysis focuses on bulk station transformers located in regions where the projected daily maximum temperature $T_{max}$ exceeds $\theta$ (°F) and where more than $\lambda$ two-day heat waves are expected per year. A substation is flagged (8) for thermal risk when both conditions are met.

$$T_{max}(S_i) > \theta^0 \ \& \ N_{hw}(S_i) > \lambda \tag{8}$$

III. RESULTS

This section presents the results of the sensitivity analysis for each climate hazard to identify thresholds beyond which mitigations are needed. The analysis was conducted using a geospatially enabled, automated framework developed in ArcGIS and Python, enabling high-resolution spatial modeling of bulk substations and associated transformers under projected climate change scenarios.

A. Exposure of Bulk Substations to Sea level Rise
1. Sensitivity Analysis

To evaluate the robustness of the coastal encroachment flagging methodology, a sensitivity analysis was conducted by varying the threshold parameter $\varphi$ (3) to flag bulk stations for further review.

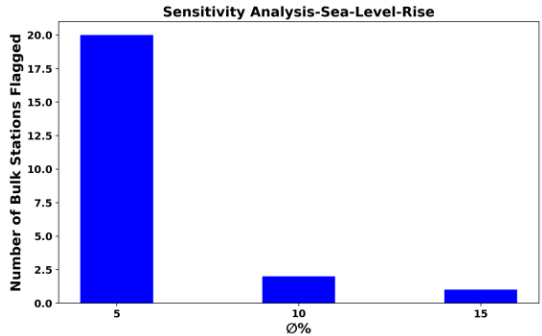

Figure 1: Number of bulk stations flagged by varying $\varphi$

The analysis shown in Fig. 1 above indicates that increasing the threshold values $\varphi$, which indicates higher confidence in the climate projections, results in an exponential decrease of the number of flagged substations. By comparing the marginal decrease of flagged stations for each level of the parameter $\varphi$, a threshold of $\varphi = 10\%$ was selected as a balanced value where marginal changes start to flatten out. At that 10% level, the following benefits are observed:

- Manageable Scope: At 10%, the number of flagged substations was reduced to a manageable level, allowing for focused engineering review and resource allocation.
- Risk Sensitivity: This threshold captures substations with meaningful proximity changes due to sea level rise, without overestimating exposure.
- Practical Relevance: A 10% inland shift often corresponds to a tangible change in coastal dynamics that could affect access, drainage, or protective barriers.
- Uncertainty Buffer: Given the uncertainties in sea level rise projections and the presence of elevation buffers at many substations, a moderate threshold helps avoid false positives while still identifying credible risks.

2. Exposed Stations Map

Geospatial analysis indicates that no substation locations are projected to experience direct inundation due to sea level rise by the year 2050. However, two substations locations, have been identified as potentially vulnerable due to a projected reduction of more than 10% in their distance to the coastal waterline relative to current conditions (Fig. 2).

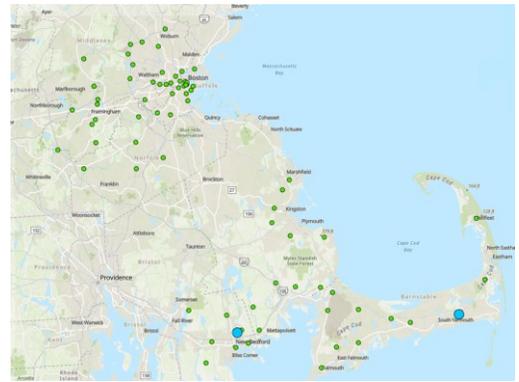

Figure 2. Bulk substations exposed to SLR are shown in blue; green indicates all substations

B. Exposure of Substations to Storm Surge

Based on the modeled storm surge exposure of a Category 1 storm, thirty (8) existing bulk substation locations are identified as potentially vulnerable to inundation (Fig. 3). These findings underscore the importance of validating substation elevation against projected surge levels and considering adaptive measures where necessary.

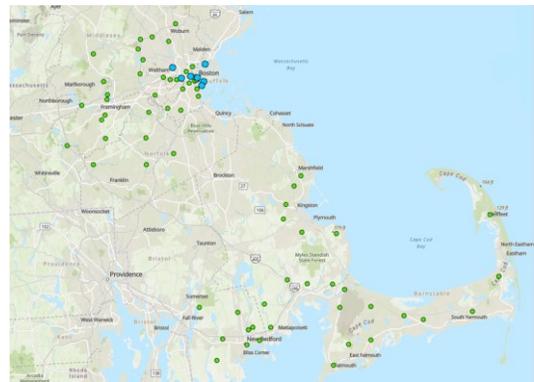

Figure 3. Bulk substations exposed to Storm Surge are shown in blue; green indicates all substations

C. Exposure of Substation to Inland Flooding
1. Sensitivity Analysis

A sensitivity analysis was performed by varying $r_s$ (radius around station to evaluate the relative elevation of the station to quantify the risk of pluvial flooding) from 0.25 miles to 1 mile and $\delta$ (distance to inland bodies of water for fluvial flooding risk) from 100 meters to 500 meters. These ranges aligns with FEMA's Risk MAP [21], which emphasizes localized flood risk assessments and infrastructure vulnerability and FEMA's Flood Insurance Rate Maps (FIRMs) [22], where flood risk is not binary but varies with distance from water bodies.

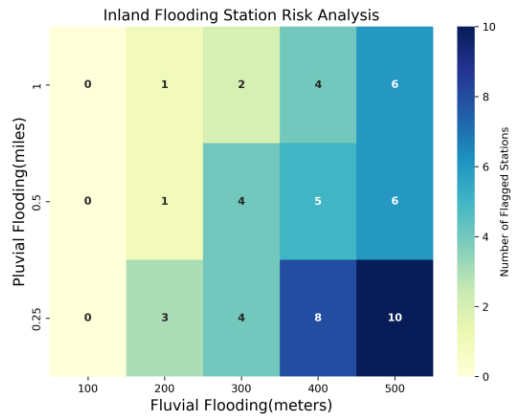

Figure 4: Sensitivity analysis of substations under different $r_s$ and $\delta$ values

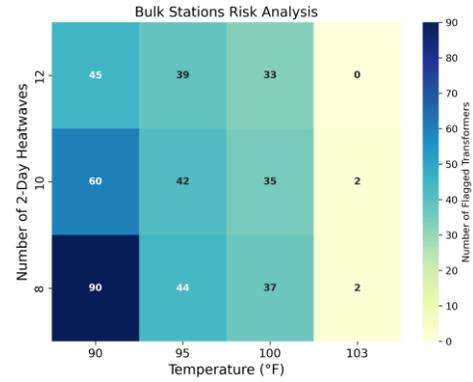

Figure 6: Sensitivity analysis of substations under different θ and λ values

The sensitivity analysis shown in Fig. 4 above indicates that reasonable thresholds are $r_s = 0.5\ miles$ and $\delta = 200\ m$. These values allow for:

- Avoiding Over-Flagging: Smaller thresholds missed substations that are realistically at risk due to terrain slope or floodplain extent.
- Balance of Local Context and Data Resolution: A 0.5-mile radius captures meaningful local topographic variation without extending into unrelated terrain features.
- Hydraulic Influence Zone: 200 meters is a reasonable approximation of the area that could be affected by riverine overflow during extreme precipitation or snowmelt events.
- Empirical Support: Preliminary testing showed that 0.5 miles radius around substation and 200 meters from water bodies consistently identified substations in topographic depressions without over-flagging.

2. Exposed Substation Mapping

There are 1 substation in Massachusetts flagged for inland flooding risk (Fig. 5).

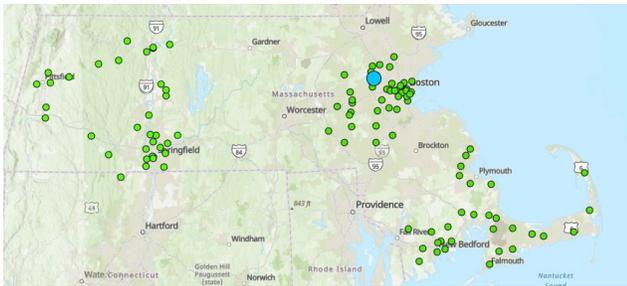

Figure 5. Bulk substations exposed to Inland Flooding are shown in blue; green indicates all substations

D. Exposure of Bulk Transformer to Warming temperature
1. Sensitivity Analysis

To evaluate the impact of threshold selection, a sensitivity analysis for SSP2-4.5 90[th] percentile scenario was conducted by varying:

- Temperature threshold, $\theta$, from 90°F to 103°F
- Heat wave frequency, $\lambda$, from 8 to 12 two-day events per year.

From above heat map (Fig 6), based on number of stations flagged, $\theta = 100$°F and $\lambda = 12$ two-day heatwave events yield a focused list of substations where adaptation strategies such as derating, enhanced cooling, or accelerated replacement can be prioritized effectively. These values were also chosen based on the following considerations:

- Climatic Relevance: 100°F represents a critical thermal threshold beyond which transformer cooling systems may become insufficient, especially under peak load conditions.
- Historical data alignment: 100°F is the highest daily temperature observed in Boston in the past ten years, 2014-2024.
- Risk Differentiation: Lower thresholds (e.g., 90°F or 8 events) flagged a large number of substations, many of which may not face immediate operational risk. Higher thresholds (e.g., 103°F or 12 events) excluded substations that are realistically vulnerable under moderate warming scenarios.
- Consistency with Utility Standards: The 100°F threshold aligns with or slightly exceeds summer emergency ratings in ANSI standards, making it a practical benchmark for reevaluation.

2. Exposed Transformers Map

Projected climate conditions under mid- and high-emissions scenarios indicate significant thermal stress on bulk station transformers by 2050. The spatial distribution of exposure where:

- Blue grids represent areas with projected average annual maximum temperatures exceeding 100°F.
- Orange grids indicate regions expected to experience more than 12 two-day heat wave events annually.
- Red grids denote areas exposed to both extreme heat conditions.
- Black dots mark the locations of potentially exposed bulk stations

Under the SSP2-4.5 90th percentile scenario (Fig. 7), approximately 33 bulk stations are projected to be exposed to both of these heat stress conditions by 2050.

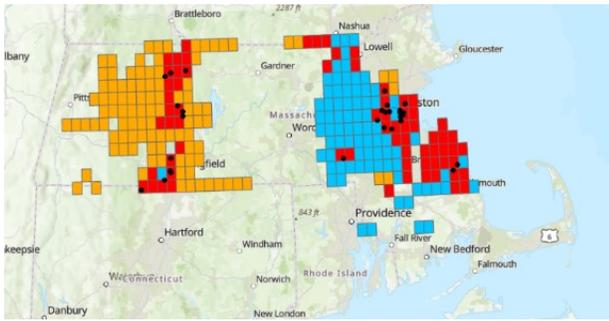

Figure 7. Bulk substations exposed to Warming Temparature under SSP2-4.5 90$^{th}$ percentile scenario

IV. CONCLUSION

This study demonstrates a practical, data-driven approach to assessing climate vulnerability in electric utility infrastructure, with a focus on substations and substation transformers. By integrating geospatial analysis, climate projections, and sensitivity thresholds, the framework enables targeted, cost-effective resilience planning. The methodology supports proactive decision-making tailored to regional conditions, offering a scalable model for utilities facing similar climate challenges. Ultimately, this work contributes to building a more resilient and adaptive electric grid in the face of evolving climate risks. Our results indicate that the number of stations requiring mitigations is highly sensitive to the thresholds for each climate hazard. This means that high-fidelity climate models are crucial to adequately hedge against climate risks and ensure a no- or least-regrets hardening approach. The use of sensitivity analysis provides a more dynamic and adaptable approach compared to conventional risk assessment methods as it provides up-to-date insights under projected 2050 scenarios and can be repeated over time to reflect evolving climate conditions.